\documentclass[twoside,a4paper,11pt]{proceedings}
\usepackage{graphicx}
\usepackage{hyperref}
\usepackage{movie15} 
\usepackage{natbib}

\begin{document}
\pagenumbering{arabic}
\pagestyle{myheadings}
\thispagestyle{empty}
\vspace*{-1cm}
{\flushleft\includegraphics[width=8cm,viewport=0 -30 200 -20]{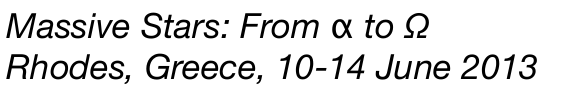}}
\vspace*{0.2cm}
\begin{flushleft}
{\bf {\LARGE
Stellar content of the young supershell GSH 305+01$-$24
}\\
\vspace*{1cm}
N. Kaltcheva$^1$,
and
V. Golev$^2$
%
}\\
\vspace*{0.5cm}
%
$^{1}$
Department of Physics and Astronomy, University of Wisconsin Oshkosh, USA \\
$^{2}$
Department of Astronomy,  St Kliment Ohridski University of Sofia, Bulgaria

\end{flushleft}
\markboth{
Stellar content of GSH 305+01$-$24
}{
Kaltcheva \& Golev et al.
}
\thispagestyle{empty}
\vspace*{0.4cm}
\begin{minipage}[l]{0.09\textwidth}
\ 
\end{minipage}
\begin{minipage}[r]{0.9\textwidth}
\vspace{1cm}
\section*{Abstract}{\small
We combine several surveys at different wavelenghts with intermediate-band $ uvby\beta$  photometry to examine the correlation between the location of the OB-stars in the Centaurus star-forming field and the neutral and ionized material in GSH 305+01$-$24 supershell seen in the same direction.
 Based on homogeneous distances of nearly 700 early-type stars, we are able to select spatially coherent stellar groupings and to revise the classical concept of the Cen OB1 association. 
We argue that this star-forming field is closer to the Sun than estimated before, at a distance of $1.8\pm0.4$ kpc, instead of the classical 2.5 kpc. 
The region shows striking similarities between the stellar distribution and H$\alpha$ and H I emission morphologies, suggesting that the observed shell passes a stage through which the number of ionizing photons emerging from the central association is not sufficient to ionize the entire shell. 
Stellar masses and ages, calculated via the latest evolutionary models are used to evaluate whether these stars can be the energy source for the supershell.
\vspace{10mm}
\normalsize}
\end{minipage}


We establish a homogeneous distance scale for nearly 700 O-B9 stars toward Centaurus. 
We  can clearly separate the intrinsically very bright, massive, recently-born stars ($M_\mathrm{V}<-2$) from
the nearby late B-type stars ($M_\mathrm{V}>-2$). 
This way we can refine the stellar content of GSH 305+01$-$24 supershell, avoiding contamination by relatively nearby and less massive early-type stars.
We detect a previously unnoticed group of very young stars at 1 kpc.
Part of the group, concentrated toward galactic coordinates ($l=305^\circ, b=1^\circ$) is heavily reddened. 
The rest of the young stars located at 1 kpc are less reddened and spread out across the field. 

The intrinsically very bright massive stars found between 1 and 3.6 kpc delineate an extended starforming field. 
According to this sample of nearly 230 massive OB stars, we are not able to confirm the classical Cen OB1 or any of the suggested subgroups of the 
association (\citep{hum78,mel95}).
From these, 133 stars seem to belong to the GSH 305+01$-$24 supershell being located at an average distance of 1.8$\pm$0.4 kpc.

A comparison of $uvby\beta$ absolute magnitudes $M_\mathrm{V}$ and MK-based $M_\mathrm{V}$ to the Hipparcos-based $M_\mathrm{V}$ points out a
possible over-estimation of stellar distance when relying on an MK based determination \citep{kal11}. 
Since the distances to the Galactic OB associations come mainly from individual stellar distances and rarely from main-sequence fitting, the $uvby\beta$
photometry should lead to a significant improvement of the distances for many OB-groups in the Milky Way.

We have used the H I radio-data (Southern Galactic Plane Survey,  \citealt{mcc05}), the 100 $\mu$m DIRBE far-IR data ( \citealt{sfd98}), 
and the H$\alpha$ data (\citealt{fin03}) to produce a multiwavelength map of the Centaurus field.
GSH 305+01$-$24 and its optical counterpart, the Coalsack Loop H$\alpha$ shell \citep{wal98}, can be associated with the massive stars around 1.8 kpc. 
The OB-group at 1 kpc is most probably connected to the nearby GSH 304$-$00$-$12 supershell at 1.2$\pm$0.8 kpc \citep{mcc01},
 but it could be also located at the near side of GSH 305+01$-$24. 

\begin{figure}
\begin{center}
\includegraphics[width=\textwidth,viewport=28 20 750 550]{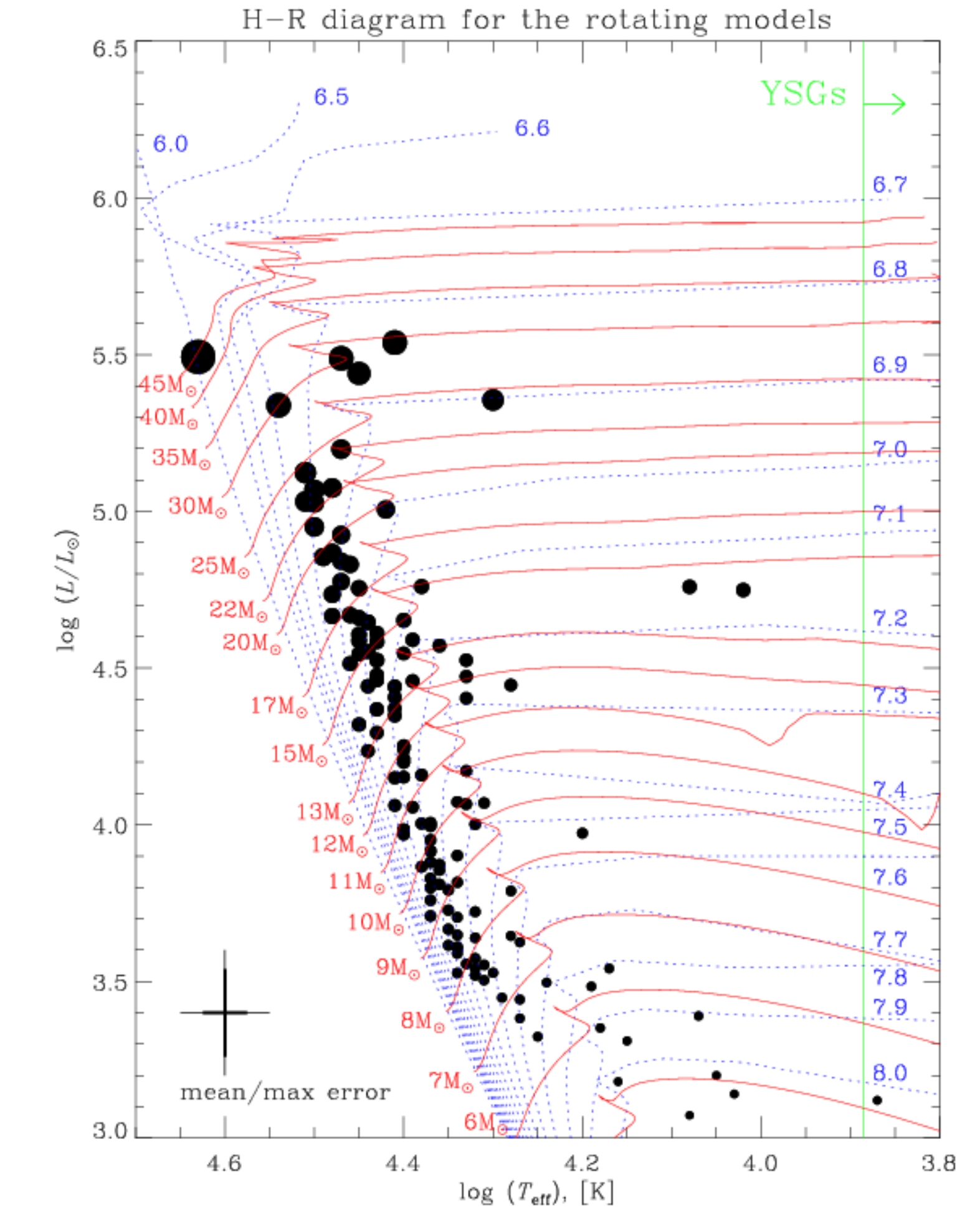}  
\caption{The HR diagram for the massive stars located inside the GSH 305+01$-$24 supershell. 
The size of the symbols corresponds to the mass of the stars. 
}
\end{center}
\end{figure}

As mentioned, we select  photometrically 133 stars at an average distance $1.8\pm0.4$ kpc located inside the GSH 305+01$-$24 supershell.
 This distance yields approximately $230\times360$ pc for the size of the shell. 
An estimate of the ages and masses of all stars via the evolutionary tracks and isochrones by \citet{eks12} results in 12.2 Myr for the average age of these stars.
This is in excellent agreement with our estimates for the distance of the young open cluster NGC 4755 (1.97 kpc, 15.1 Myr), also located inside the shell.
In Fig.~1 the HR diagram for the massive stars inside the GSH 305+01$-$24 is shown. 
Models with initial metallicity 0.014 (solar) and initial rotation rate 0.40 are plotted.

The morphology of H$\alpha$ shell traces the stellar distribution, as in a wind-driven H$\alpha$ shell.
For each star the size of the its wind-blown bubble is estimated from the relation between the bubble size and
the initial mass of the star reported by \citet{che13}. 
The cumulative effect of the stellar winds of the selected stars and NGC 4755 is sufficient to produce the observed size of the H I shell. 
Our results, based on revised stellar content and distance to GSH 305+01$-$24, are in agreement with the conclusions of \citet{mcc01} 
and the models of \citet{sil08}  that the GSH 305+01$-$24 is a wind-blown shell. 

\small  
%
\section*{Acknowledgments}   
%

This research has made use the SIMBAD database, operated at CDS, Strasbourg, France, and 
NASA's SkyView Virtual Observatory (http://skyview.gsfc.nasa.gov), a service of the 	
Astrophysics Science Division at NASA/GSFC and the High Energy Astrophysics Division of the SAO.
This research is supported by the National Science Foundation grant AST-0708950.

\bibliographystyle{aj}
\small
\bibliography{NKaltcheva}

\end{document}